\documentclass[fleqn,usenatbib]{mnras}

\usepackage{newtxtext,newtxmath}

\usepackage[T1]{fontenc}

\DeclareRobustCommand{\VAN}[3]{#2}
\let\VANthebibliography\thebibliography
\def\thebibliography{\DeclareRobustCommand{\VAN}[3]{##3}\VANthebibliography}

\usepackage{graphicx}	
\usepackage{amsmath}	

\def \src {XTE~J1906+090}
\def \swift {\emph{Swift}}
\def \int {$INTEGRAL$}
\def \xmm {$XMM$-$Newton$}

\def \sw {$Swift$}
\def \xte {$RXTE$}
\def \chandra {$Chandra$}

\def \hcm {\hbox {\ifmmode $ atom cm$^{-2}\else atom cm$^{-2}$\fi}}

\def \arcsec {\hbox{$^{\prime\prime}$}}

\def \apjl {ApJL}
\def \apjs {ApJS}
\def \aap {A\&A}

\title[]{XTE J1906+090: a persistent low luminosity Be X-ray Binary}

\author[Sguera et al.]
{V. Sguera$^{1}$, L. Sidoli$^2$,  A. J. Bird$^3$,  N. La Palombara$^2$ \\ 
$^1$  INAF$-$OAS, Osservatorio di Astrofisica e Scienza dello Spazio, Area della Ricerca del CNR, via Gobetti 101, I-1-40129 Bologna, Italy \\
$^2$ INAF$-$IASF, Istituto di Astrofisica Spaziale e Fisica Cosmica, Via A. Corti   12, I-20133 Milano, Italy \\
$^3$ School of Physics and Astronomy, Faculty of Physical Sciences and Engineering, University of Southampton, SO17 1BJ, UK \\
}

\date{Accepted 2023 May 8. Received  in original form 2023 March 17.}

\pubyear{2023}

\begin{document}
\label{firstpage}
\pagerange{\pageref{firstpage}--\pageref{lastpage}}
\maketitle

\begin{abstract}
We present new results from \int\ and \sw\ observations of the hitherto poorly studied and unidentified X-ray source XTE~J1906+090. A bright hard X-ray outburst (luminosity of $\sim$10$^{36}$ erg s$^{-1}$ above 20 keV) has  been discovered with  \int\  observations in 2010, this being the fourth outburst ever detected from the source.  Such  events are sporadic, the source duty cycle is in the range (0.8--1.6)\%  as  inferred from extensive \int\ and \sw\  monitoring in a similar hard X-ray  band. Using five archival unpublished \sw/XRT observations, we found that \src\  has been consistently detected at a persistent low X-ray luminosity value of $\sim$10$^{34}$ erg s$^{-1}$, with  limited variability (a factor as high as  4). Based on our findings, we propose that \src\ belongs to the  small and rare group  of  persistent low luminosity Be X-ray Binaries. 
\end{abstract}




\section{Introduction}
XTE~J1906+090 is an 89 seconds transient X-ray pulsar serendipitously discovered  by \xte\ in 1996 
(Marsden et al. 1998). 
To date only two outbursts have been reported in the literature, detected  in 1996 and 1998  with RXTE/PCA during regular a monitoring campaign of the nearby magnetar
SGR J1900+14 (Wilson et al. 2002). The two outbursts differed significantly in term of peak flux and duration. The 1996 outburst  was particularly short (2-4 days duration) and not particulary bright since it reached a peak flux  of  $\sim$ 6$\times$10$^{-12}$ erg cm$^{-2}$ s$^{-1}$  (2--10 keV). The 1998 outburst was longer (25 days duration) and brighter (2-10 keV peak flux of  $\sim$ 1.1$\times$10$^{-10}$ erg cm$^{-2}$ s$^{-1}$). From spectral analysis, the measured high absorption column density suggested a source  distance  of at least 10 kpc. 
\chandra\ observations in 2003 (G{\"o}{\v{g}}{\"u}{\c{s}} et al. 2005) caught the source at a flux level of $\sim$ 3$\times$10$^{-12}$ erg cm$^{-2}$ s$^{-1}$ (2-10 keV). The arcsecond sized \chandra\ position allowed  pinpointing the optical/infrared counterpart whose colors are typical of an early type star  (G{\"o}{\v{g}}{\"u}{\c{s}} et al. 2005). Based on the above evidence, XTE J1906+090 has been considered a candidate transient Be X-ray Binary (BeXRB). However  the firm confirmation of such nature can only be obtained through optical/infrared spectroscopy which is still lacking. Corbet et al. (2017) performed a \swift/BAT investigation of the long term hard X-ray light curve of the source. They noted the presence of two bright outbursts in the 15-50 keV light curve  (dated 2009 and 2016), although no information was provided about their  duration and energetic. In addition,  Corbet et al. (2017) found two periodicities at  $\sim$81 days and $\sim$173 days. Although their nature is still unclear,  they might be the orbital and super-orbital period, respectively.     

Here we report new results on \src\  obtained from unpublished  \swift/XRT and   \int\ archival data.

\section{Data Analysis}

\subsection{INTEGRAL}

The temporal and spectral behaviour of XTE~J1906+090 has been  investigated  in detail above 20 keV with  the ISGRI detector (Lebrun et al. 2003), which is the lower energy layer of the IBIS coded mask telescope (Ubertini et al. 2003) onboard \int~(Winkler et al. 2003).

 \emph{INTEGRAL} observations are divided into short pointings (Science Windows, ScWs) whose 
typical duration is $\sim$ 2,000 seconds. The IBIS/ISGRI public data archive (from revolution 30 to 2,000, i.e. from approximately January 2003  to September 2018)  
has been specifically searched for hard X-ray activity from  \src. In particular, the data set consists of
8,119 ScWs where  \src\ was  within the instrument  field of view (FoV) with an off-axis angle value less than 12$^{\circ}$. 
We note that such a limit  is generally applied because the
response of IBIS/ISGRI is not well modelled at large off-axis values and this  may introduce a systematic error in
the measurement of the source fluxes.  

To search for transient hard X-ray  activity from \src~in a systematic way, we used the bursticity method as developed by 
Bird  et al. (2010, 2016). Such method optimizes the source detection time-scale by scanning the IBIS/ISGRI light curve with a variable-sized  time window to search for the best source significance value. Then the duration, time interval and energy band  
over which the source significance is maximized,  are  recorded. Once a newly discovered outburst activity from \src~was found, a specific spectrum and light curve were  extracted. To this aim, the data reduction was carried out with the latest release 11.2 of the Offline Scientific Analysis software (OSA, Courvoisier et al. 2003).  The IBIS/ISGRI systematics, which  are typically of the order of 1$\%$,  were added  to the extracted spectrum.  

Images from the X-ray Monitor JEM--X  (Lund et al. 2003) onboard \int\  were created in two energy bands (3--10 and 10--20 keV) for  the 
outburst reported later in this work.

\subsection{The Neil Gehrels \swift\ Observatory}
\label{sect:swiftobs}

The Neil Gehrels \swift\ Observatory (\swift\ hereafter; Gehrels et al. 2004) 
has covered the source sky position with the
X-ray Telescope (XRT;  Burrows et al. 2005) six times (Table~\ref{tab:swift_log}).
The first three observations were nominally pointed at \src, while the other three imaged
the source position serendipitously, at very large off-axis angle.

We have reprocessed these XRT data (operated in photon counting (PC) mode)
with {\tt HEASoft} v6.29 and the most updated calibration files,
using the {\tt XRTPIPELINE} tool with standard procedures.
Source detection and estimate of the source net count rates 
have been obtained using {\tt XIMAGE} (v4.5.1) in {\tt HEASoft} and the 
{\tt SOSTA} and {\tt UPLIMIT} tools, when an upper limit to the source count rate was needed.

Once all the single XRT snapshots have been reprocessed (and the ancillary response files, ARF, built with {\tt XRTPIPELINE}), 
we merged together the three observations 
performed in 2018, showing the lowest offaxis angle (Table~\ref{tab:swift_log}), for a meaningful spectroscopy (see section 3.2).
A single XRT spectrum has been extracted from the merged event file, 
using a circular region with a radius of 20 pixels (i.e. $\sim$47\arcsec) centered at the source position. 
The background spectrum was extracted from an annular region around the source centroid, 
adopting radii of 30 and 50 pixels for the inner and outer radii,  respectively.
The appropriate, weighted average ARF has been produced merging 
together the three single ARF files, using the {\tt ADDARF} tool.
The source spectrum has been grouped to have at least one count per bin and then fitted  
using {\tt XSPEC} v12.12.0 (Arnaud 1996), adopting Cash statistics (Cash1979). 

The uncertainties on the count rates are computed at 1$\sigma$ confidence level, 
while the uncertainties on spectral parameters and fluxes
calculated in {\tt xspec} are at 90\% confidence level.
All upper limits are calculated at 3$\sigma$ level.
X-ray fluxes have been estimated in the 0.3-10 keV range, with  
uncertainties calculated using {\tt cflux} in {\tt xspec}.
The model {\tt TBabs} has been adopted to account for the absorbing column density along the line of sight, 
assuming the interstellar abundances of (Wilms et al. 2000) and photoelectric absorption cross sections of 
(Verner  et al.1996).

\begin{table}
	\centering
	\caption{Summary of the \swift\ observations}
	\label{tab:swift_log}
	\begin{tabular}{lccc} 
		\hline
		\hline
Obs ID & start time             & exp  &   off-axis angle \\
               & (UTC)                  & (ks) &  (arcmin) \\	
		\hline
   00010740001 &  2018-07-04 17:36:00   & 0.67   & 4.6 \\
   00010740002 &  2018-07-08 08:46:00   & 1.16   & 1.9 \\
   00010740003 &  2018-07-10 11:56:00   & 1.60   & 1.8 \\
   00014213001 &  2021-03-31 01:53:35   & 1.30    &  12.1  \\
   00014213002 &  2021-04-14 13:09:34   & 0.16    &  12.0   \\
   00014213004 &  2021-04-28 16:24:34   & 1.80    &  11.9   \\
		\hline
	\end{tabular}
\end{table}

\section{Results}
\label{sect:res}

\subsection{INTEGRAL}

XTE J1906+090 has been best detected with the bursticity method in the energy band 18--60 keV during outburst activity on September 2010. Here we report, for the first time, on a detailed temporal and spectral analysis of this hard X-ray outburst.  This is the first activity from the source reported since \xte\ detections during the late nineties.  

As we can note from Table 2, XTE~J1906+090  was not detected in  the single revolutions 965 (9 Sep 2010) and 966 (11 Sep 2010), nor in their summed significance map for a total on source exposure of $\sim$ 10 ks.  We inferred a 2$\sigma$ upper limit of $\sim$ 3 mCrab (18--60 keV).
Conversely, the  source was detected  with a significance of  6.5$\sigma$ and 6.0$\sigma$ (18--60 keV) during revolution 972 (29 Sep 2010) and 973 (2 Oct 2010), respectively. We stacked the data for  both revolutions with the aim of increasing the statistics.  XTE~J1906+06 was detected  with a significance of 8.5$\sigma$ (18--60 keV) for a total on-source exposure of $\sim$ 10.5 ks (see Fig. 1). No detection was obtained in the higher energy band 60--100 keV. The average 18--60 keV flux is 
12.9$\pm$1.5 mCrab (or $\sim$ 1.7$\times$10$^{-10}$ erg cm $^{-2}$ s$^{-1}$). The source was never significantly detected 
 at ScW level  (i.e. $\ge$ 5$\sigma$) at any point  of the observation. Finally,  XTE~J1906+090  was not detected in any single revolution after 973, i.e. 974 (7 Oct 2010) and 976 (10 Oct 2010), nor in their summed mosaic for a total on source exposure of $\sim$ 9 ks.  We inferred a 2$\sigma$ upper limit of $\sim$ 3.3 mCrab (18--60 keV).

  \begin{table}
\caption {List of  {\itshape INTEGRAL}  orbits and their start date,  significance of source detection, average flux (18--60 keV) and effective exposure time on-source.} 
\label{tab:main_outbursts} 
\begin{tabular}{ccccc}
\hline
\hline    
Orbit    &     start time                                 &    Significance            &    Flux                &  exposure       \\
    (n.)          &  (MJD)                           &   ($\sigma$)                     &    (mCrab)            & (ks)  \\
\hline  
965+966           &       55450.088       &         $<$2$\sigma$           &    $\le$ 3.1               &     10.0  \\  
972+973                   &      55468.587       &        8.5$\sigma$            &  12.9$\pm$1.5        & 10.5       \\  
974+976           &      55476.263         &        $<$2$\sigma$               &    $\le$ 3.3              &           8.8 \\
\hline
\end{tabular}
\end{table}


\begin{figure}
\begin{center}
\includegraphics[height=8 cm, angle=0]{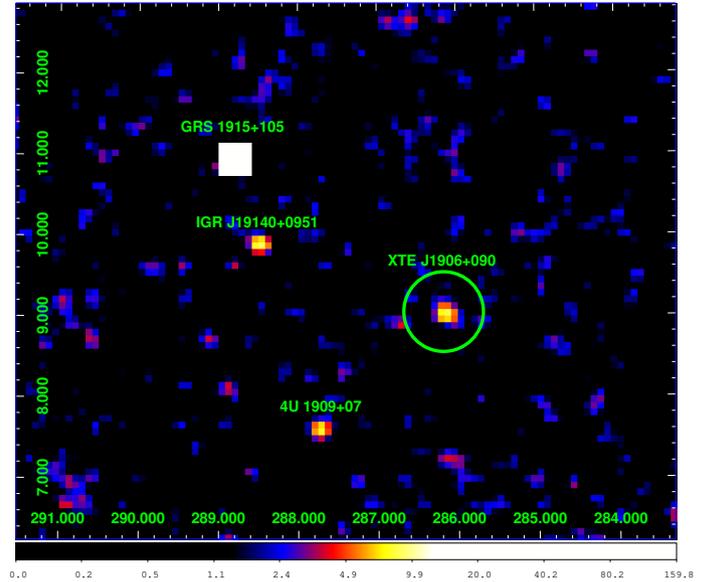}
\caption{IBIS/ISGRI 18--60 keV significance map (revolutions 972+973), XTE~J1906+090 (green circle) is detected at 8.5$\sigma$ level.}
\end{center}
\end{figure}

\begin{figure}
\begin{center}
\includegraphics[height=6.5 cm, angle=0]{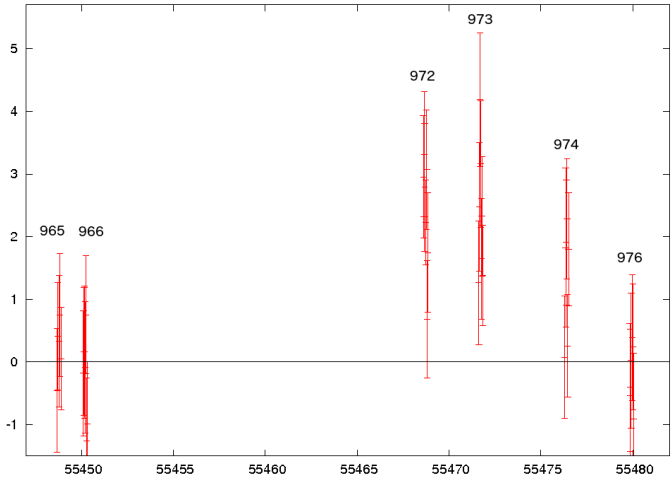}
\caption{IBIS/ISGRI light curve of  XTE~J1906+090 (18--60 keV) at ScW level (2,000 s) covering the INTEGRAL observations listed in Table 2. Time and flux axis are in MJD and count s$^{-1}$, respectively.}
\end{center}
\end{figure} 
 
 The IBIS/ISGRI light curve at ScW level (Fig. 2) clearly shows the outburst activity detected during revolutions 972 and 973. It was characterized by a duration of $\sim$ 
 4 days  during which the measured peak flux (at MJD 55471.7) is  21$\pm$5 mCrab or  $\sim$ 2.7$\times$10$^{-10}$ erg cm$^{-2}$ s$^{-1}$  (18--60 keV),
 Unfortunately, there is a gap in the light curve between revolutions 966 and 972 which does not allow us to fully constrain the proper duration of the total outburst activity.  
 Based on the \emph{INTEGRAL} data coverage,  the  duration can be  loosely constrained  in the range 3--25 days. 
In this context, \emph{Swift/BAT} data can be particularly helpful since the BAT instrument provides a continuous time coverage of the source activity with no temporal gaps. From the \swift/BAT archive,  we downloaded the 15--50 keV source light curve  on daily  time-scale, and we checked if the source was active or not 
during the temporal range corresponding to the gaps of the IBIS/ISGRI light curve. No significant hard X-ray activity was evident, suggesting that the duration of the outburst  detected by \int\ was  very likely of the order of $\sim$ 4 days. Such a  duration is similar to that estimated with \xte\ during the outburst occurred in 1996.

We extracted the average IBIS/ISGRI spectrum of the outburst during revolutions 972+973.  
The best fit is achieved with a power law model ($\chi^{2}_{\nu}$=0.92, 4 d.o.f.) characterized by a steep photon index ($\Gamma$=4.7$\pm$0.7). 
The average 18--60 keV (20--40 keV) flux is 1.2$\times$10$^{-10}$ erg cm$^{-2}$ s$^{-1}$ (8.3$\times$10$^{-11}$ erg cm$^{-2}$ s$^{-1}$). 
Alternatively, a  thermal bremsstrahlung model provides a reasonable fit  ($\chi^{2}_{\nu}$=0.6, 4 d.o.f.) with   kT=7.7$^{+2.7}_{-1.8}$ keV. 
Fig. 3  shows the power law data-to-model fit with the corresponding residuals.

The source was also in the JEM--X2 FoV during  INTEGRAL observations  from revolution 972 to 973, for a total on-source exposure of  $\sim$15 ks. 
No  detection was achieved in the X-ray band 3--10 keV, the inferred 3$\sigma$ upper limit  is of the order of $\sim$ 6 mCrab or 9$\times$10$^{-11}$ erg cm $^{-2}$ s$^{-1}$.  In the higher energy band 10--20 keV there is a hint of weak detection at  $\sim$4$\sigma$ level. The statistics are insufficient to perform any meaningful analysis. Eventually we do not consider it  as a reliable detection. 

 XTE~J1906+090 is not detected as a persistent source in the latest \emph{INTEGRAL/IBIS} catalog of Bird et al. (2016), despite an extensive coverage of its sky 
region for a total on-source exposure of $\sim$ 6 Ms.  This information can be used to infer a 3$\sigma$ upper limit on its persistent hard X-ray emission, which  is 
of the order of 0.3 mCrab or 2.3$\times$10$^{-12}$ erg cm$^{-2}$ s$^{-1}$ (20--40 keV). When assuming the source flux in the same energy band 
as measured by IBIS/ISGRI during  the reported outburst, we can derive a dynamic range $\ge$ 36.

\begin{figure}
\begin{center}
\includegraphics[height=5.9 cm, angle=0]{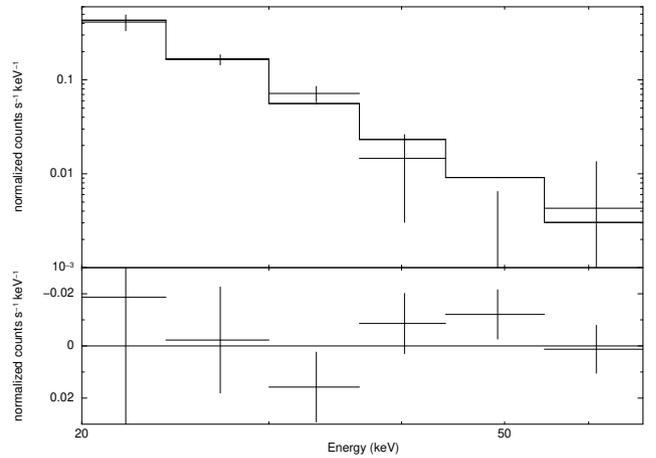}
\caption{IBIS/ISGRI spectrum  of XTE~J1906+090, showing the power law data-to-model fit with the corresponding residuals.}
\end{center}
\end{figure}

\subsubsection{Search for periodicities}
We performed a timing analysis of the long term IBIS/ISGRI light curves, 
using the epoch folding method, in order to search for periodicities which could eventually be ascribed to the orbital or super-orbital period as previously found by Corbet et al. (2017). 
The IBIS/ISGRI ScW light curve (18-60 keV) covers the observational period from 2003 Jan  to 2018 Sep. We applied standard optimum filtering in order to exclude poor quality data points which could eventually disrupt the periodic signal (see Goossens et al. 2013 and Sguera et al. 2007 for details). Periodicities were searched for in the range  0.1-1000 days, but none were found.

\subsection{\swift}
\label{sect:swift_res}

A source positionally coincident with \src\ has  been detected in each single XRT pointing,
except in ObsID~00014213002,
where the very short exposure time allowed us to obtain only an upper limit to the source intensity. 
The source count rates (corrected for PSF, sampling dead time and vignetting) 
obtained using {\tt SOSTA} and {\tt UPLIMIT} tool in {\tt XIMAGE} on the  
single images and exposure maps are reported in Table~\ref{tab:xrtrates}.
We have converted these XRT/PC rates into unabsorbed fluxes (0.3-10 keV),
adopting an average factor of 1.$\times$10$^{-10}$ erg cm$^{-2}$ count$^{-1}$
(assuming a power law model with a photon index $\Gamma$=2 and N$_{\rm H}=1\times10^{22}$cm$^{-2}$).

\begin{figure}
 \includegraphics[width=9cm, angle=0]{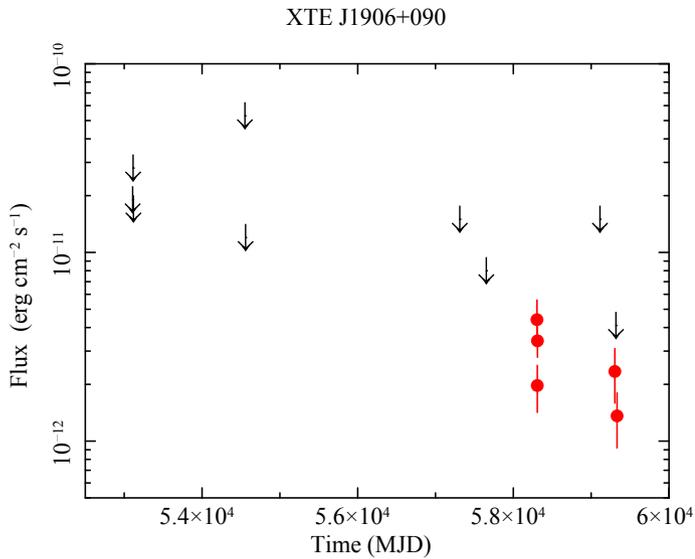} 
 \caption{Long-term light curve of \src\  (\swift-XRT and \xmm\ data). 
The unabsorbed \swift-XRT fluxes are derived from Table~\ref{tab:xrtrates} (see text), while 
the 3$\sigma$ upper limits on fluxes obtained from the \xmm\ slew observations
are taken from Table~\ref{tab:xmmslew}. 
A power law model with a photon index $\Gamma$=2 and N$_{\rm H}$=10$^{22}$cm$^{-2}$ has been assumed.
}
\label{fig:lc}
\end{figure}

In Fig.~\ref{fig:lc} we show the long-term light curve built from these fluxes,
together with several upper limits derived from \xmm\ Slew Observations (see Sect.~\ref{sect:xmmslew}).

\begin{table}
	\centering
	\caption{\src\ intensity during the \swift\ XRT observations}
	\label{tab:xrtrates}
	\begin{tabular}{lcc} 
		\hline
		\hline
Obs ID & start date              & Count rate       \\
               & (MJD)                   & (10$^{-2}$ count s$^{-1}$) \\	
		\hline
   00010740001 &  58303.7333             &   4.4$\pm{1.2}$      \\         	
   00010740002 &  58307.3652             &   1.97$\pm{0.55}$    \\
   00010740003 &  58309.4972             &   3.4$\pm{0.62}$     \\
   00014213001 &  59304.0789             &   2.34$\pm{0.75}$    \\
   00014213002 &  59318.5483             &   $<$4.1           \\
   00014213004 &  59332.6837             &   1.36$\pm{0.44}$      \\
		\hline
	\end{tabular}
\end{table}

\begin{figure}
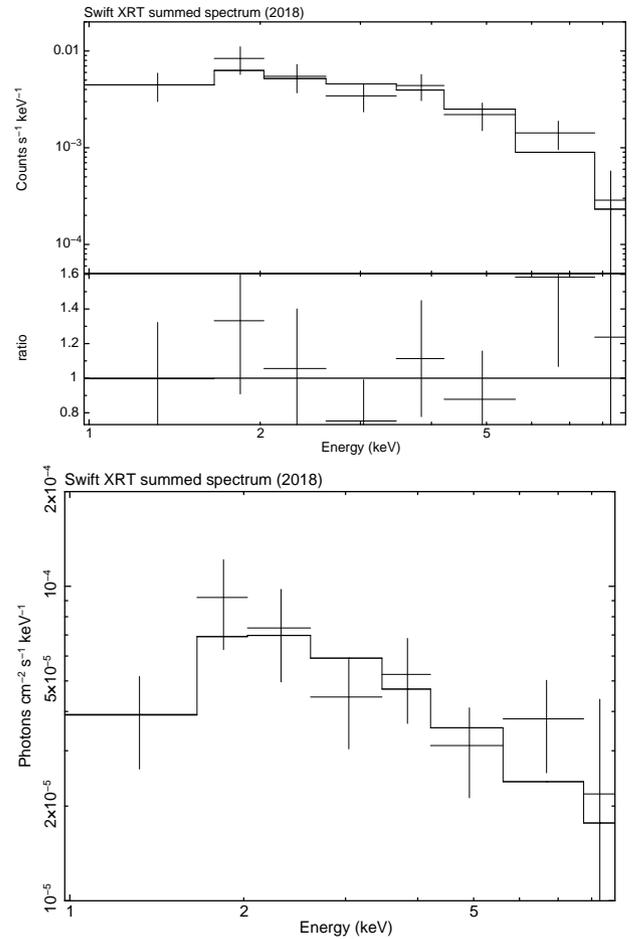

 \includegraphics[width=6cm, angle=270]{fig5a.eps} 
  \includegraphics[width=6.3cm, angle=270]{fig5b.eps}
\caption{\swift-XRT spectrum obtained from the merging of the three observations performed in 2018. 
An absorbed power law model is shown (see text for the parameters): the counts spectrum together with
the ratio with respect to the best-fit model is shown in the \textit{top panel}, while the 
photon spectrum is displayed in the \textit{bottom panel}. 
Spectra have been rebinned for presentation purposes, only.
}
    \label{fig:swift_spec}
\end{figure}

We focussed our attention on the three contiguous XRT observations which spans the time range  4-10  July 2018 (see Table 1). Since the XRT spectra extracted  from each single observation have a very similar spectral shape, we considered the spectrum extracted from their sum. This spectrum resulted in  a net count rate of 0.022$\pm{0.003}$~count~s$^{-1}$ (0.3-10 keV), for an exposure time of 3.4 ks. 
A good fit has been obtained with an absorbed power law model (C-stat 57.93 using 74 bins;
null hypothesis probability of 0.981 with 71 degrees of freedom, dof), with the following
spectral parameters:
absorbing column density NH=2.0($^{+2.0} _{-1.5}$)$\times$10$^{22}$~cm$^{-2}$, 
power law photon index $\Gamma$=1.38$^{+0.95}_{-0.83}$, and
a flux corrected for the absorption UF=3.2($^{+3.4} _{-0.9}$)$\times$10$^{-12}$~erg~cm$^{-2}$~s$^{-1}$.
The spectrum is shown in Fig.~\ref{fig:swift_spec}. We note that these spectral shape and flux values are very similar to those 
measured during the 2003  \chandra\  observation reported by G{\"o}{\v{g}}{\"u}{\c{s}} et al. (2005).

%

\subsection{\xmm\ Slew Observations}
\label{sect:xmmslew}

The source sky position has never been covered by any \xmm\  (Jansen et al. 2001) pointing.
However, a few slew manoeuvres (which reoriented the spacecraft between targets)
serendipitously covered \src\ position with very short exposure times.
We decided to use  these \xmm\ slew observations to constrain the long-term source light curve. 
To this aim, we have used the ESA upper limit server 
(HIgh energy LIghtcurve Generator,
HILIGT\footnote{http://xmmuls.esac.esa.int/upperlimitserver/}; Saxton et al. (2022), K{\"o}nig et al. (2022), 
which calculates  the 3$\sigma$ upper limits on the X-ray fluxes from
the \xmm\ slew observations.
The results are reported in Table~\ref{tab:xmmslew}.

We note that by default, HILIGT returns 3$\sigma$ upper limits on absorbed fluxes (0.2-12 keV) assuming a power law model with
a photon index, $\Gamma$, of 2 and a low absorbing column density N$_{\rm H} = 1\times10^{21}$cm$^{-2}$ (forth column in Table 4).
Therefore, we have converted these values to the 0.3-10 keV energy band (for a more direct comparison with \swift\ results), 
assuming a more appropriate column density of N$_{\rm H} = 1\times10^{22}$cm$^{-2}$ and a power law model with $\Gamma$=2.
The 3$\sigma$ upper limits on the fluxes corrected for the absorption (0.3-10 keV) are listed in the last column of Table~\ref{tab:xmmslew}, and reported in the long-term light curve (Fig.~\ref{fig:lc}).

All the  \xmm\ unabsorbed flux values reported in Table 4  imply 3$\sigma$ upper limits on the X-ray luminosity in the range from 
8.1$\times$10$^{34}$~erg~s$^{-1}$ to 5.6$\times$10$^{35}$~erg~s$^{-1}$ (see next section for the assumed distance value).

\begin{center}
\begin{table}
{\small 
\caption{3$\sigma$ upper limits on the X-ray fluxes obtained through HILIGT 
(HIgh energy LIghtcurve Generator) with $XMM-Newton$ slew observations.}
\label{tab:xmmslew}
\hfill{}
\begin{tabular}{l l c l l}
\hline\hline
 Date       & MJD           & Exp.     &  Abs.Flux$^{a}$     & Unabs.Flux$^{b}$      \\   
            &               & (s)      &  (0.2-12 keV)        &  (0.3-10 keV)   \\
\hline   
 2004-04-13 & 53108.8729   & 9.3       &  $<6.0\times10^{-12}$         &  $<1.9\times10^{-11}$ \\       
 2004-04-19 & 53114.8875   & 4.5       &  $<8.9\times10^{-12}$         &  $<2.8\times10^{-11}$ \\    	
 2004-04-23 & 53118.7965   & 6.0       &  $<5.4\times10^{-12}$         &  $<1.7\times10^{-11}$ \\       
 2008-03-25 & 54550.9285   & 1.4       &  $<16.7\times10^{-12}$        &  $<5.3\times10^{-11}$ \\       
 2008-04-03 & 54559.2683   & 9.5       &  $<3.8\times10^{-12}$         &  $<1.2\times10^{-11}$ \\       
 2015-10-16 & 57311.5233   & 6.2       &  $<4.7\times10^{-12}$         &  $<1.5\times10^{-11}$ \\       
 2016-09-21 & 57652.5307   & 9.8       &  $<2.6\times10^{-12}$         &  $<0.8\times10^{-11}$ \\       
 2020-09-20 & 59112.6833   & 8.7       &  $<4.6\times10^{-12}$         &  $<1.5\times10^{-11}$ \\       
\hline
\end{tabular}}
\hfill{}
\\
\\
$^{a}$ Absorbed flux, for an assumed power law model ($\Gamma$=2) and absorption N$_{\rm H} = 1\times10^{21}$cm$^{-2}$, in the energy band 0.2-12 keV.\\
$^{b}$ Unabsorbed flux,  calculated assuming a power law model ($\Gamma$=2) 
and absorption N$_{\rm H} = 1\times10^{22}$cm$^{-2}$, in the energy band 0.3-10 keV.\\
\end{table}
\end{center}

\section{source distance and X-ray luminosities}

Precise information on the source distance is necessary to  calculate the correct associated luminosities. To date,  the distance of XTE J1906+090 has been assumed to be of the order of 10 kpc, as merely suggested from  the measured high absorption column density obtained through \xte\ spectral  analysis (Marsden et al. 1998). 
Nowadays accurate and reliable distance estimates are  available from the GAIA astrometric mission. 
We relied on the last release available EDR3  (Gaia Collaboration et al. 2021), and on source distances computed by 
Bailer-Jones et al. (2021)  who used the parallaxes along with parameters that affect the parallax
measurement such as the magnitude and the color of the star, and a distance prior based on the density of stars in the galaxy.

Two distance estimates are available for XTE J1906+090 obtained with two different methods: geometric d$_{g}$=6.37$^{+3.47}_{-1.97}$ kpc  (based  only
on the parallax) and photogeometric d$_{pg}$=9.38$^{+2.90}_{-1.77}$ kpc  (which uses also the color and the apparent magnitude of the star).  
On the one side, the nominal values of the distance from the two methods differ by a factor of  $\sim$ 1.5, accordingly the relative luminosity values would differ by a factor  $\sim$ 2.2. On the other side, in both cases the formal errors are relatively large and  the corresponding uncertainty ranges are partially compatible. 
As for the choice of whether to use geometric or photogeometric distance, we note that for stars with negative parallaxes and large parallax uncertainties (which is the case of our specific source) photogeometric distances will generally be more precise and reliable than geometric ones (Bailer-Jones 2021). Therefore, according to the GAIA results, we will adopt a distance of 9.4 kpc  to calculate the luminosities of  \src. This is also compatible with a previous estimate from X-ray spectral analysis (Marsden et al. 1998).

In Table 5 we report the X-ray luminosities   corresponding to  the  \int\  and \swift/XRT  detections reported for the first time in this work.  In addition,   we  also report the \chandra, \xte\ and \swift/BAT X-ray luminosities corresponding to the five  detections  previously reported in the literature.

\begin{center}
\begin{table}
{\small
\caption{Summary of average (unless specified otherwise) X-ray luminosity values of all  X-ray detections  from \src\  reported to date, along with information on date of beginning activity, X-ray band and reference of the detection}
\label{tab:lum}
\hfill{}
\begin{tabular}{l  c c c c}
\hline\hline
 Satellite            & Date                      &   Lum                                           &  energy      & ref             \\   
                         &  (MJD)                   &   (erg s$^{-1}$)                           &    (keV)                          &         \\
\hline   
RXTE             & 50343.000         &  6.3$\times10^{34}$ $\ast$   &     2-10        &  (1),(2) \\  
RXTE             & 51056.000         &   1.1$\times10^{36}$  $\ast$         &     2-10        &  (2)  \\  
Chandra         & 52785.000         &  3.1$\times10^{34}$          &     2-10        &  (3) \\  
Swift/BAT       & $\sim$55000               & 3.6$\times10^{36}$  $\ast$    &     18-60     & (4) \\  
INTEGRAL      & 55468.587            &  3.0$\times10^{36}$   $\ast$       &     18-60     & this work \\ 
Swift/BAT       & $\sim$57450             & 3.1$\times10^{36}$ $\ast$       &     18-60     & (4) \\  
Swift/XRT               & 58303.733            & 4.6$\times10^{34}$         &     0.3-10     & this work \\   
Swift/XRT               & 58307.365            & 2.1$\times10^{34}$          &     0.3-10     & this work \\  
Swift/XRT               & 58309.497           & 3.6$\times10^{34}$         &     0.3-10     & this work \\  
Swift/XRT         & 59304.907         & 2.5$\times10^{34}$          &     0.3-10     & this work \\  
Swift/XRT          & 59332.683          & 1.5$\times10^{34}$          &     0.3-10     & this work \\  
\hline
\end{tabular}}
\hfill{}
\\
\\
(1) Marsden et al. 1998; (2) Wilson et al. 2002;  (3) G{\"o}{\v{g}}{\"u}{\c{s}} et al. 2005; (4) Corbet et al. (2017); 
$\ast$ = peak luminosity

\end{table}
\end{center}

\section{Discussion}

\src\ is an unidentified transient X-ray pulsar whose  optical/infrared counterpart  has colors typical of an early type star. Hence it is considered a candidate  BeXRB,  although  spectroscopy  is still lacking.

Table 5  lists a summary of  characteristics of all  X-ray detections  of \src\  obtained to date, the majority of them   are  reported in this work  for the first time.  We note that four X-ray outbursts have been detected, thanks to an extensive  source monitoring spanning  over three decades with X-ray missions like \xte,  \int\ and \swift. Moreover,  we note that the \xte\ 1996 detection (MJD 50343)  has been previously reported in the literature as an X-ray outburst by Wilson et al. (2002), however we argue that its peak-luminosity  (6$\times$10$^{34}$ erg  s$^{-1}$) is more than one order of magnitude lower than typical Type I outbursts from BeXRBs.  We are more inclined to consider it as enhanced X-ray activity  due to accretion of material from the stellar wind of the Be star and not from its decretion disk.    

The four outbursts are characterized by a similar X-ray luminosity of $\sim$10$^{36}$ erg  s$^{-1}$.    The first outburst  was detected in 1998 by \xte\   (2-10 keV) with a duration of $\sim$ 25 days (Wilson et al. 2002). The  outburst  detected by  \int\ (18-60 keV) is newly discovered  and reported  in this work for the first time; it was detected  in 2010 with a duration of  $\sim$ 4 days.  
According to Corbet et al. (2017),  we considered $\phi$=0 at MJD 55,000 and the  periodicity of 81.4 days (173.1 days)  to measure the  phase of the \int\ outburst.  We found that  its peak  took place at  $\phi$$\sim$0.8 ($\phi$$\sim$0.7). For a comparison, we refer the reader to  Fig. 20 and 21 in Corbet et al. (2017) where  the  light curve  folded on 81.4 days (173.1 days) clearly shows  a smooth  (sharp) peak at $\phi$$\sim$0.9-1 ($\phi$$\sim$1).
Two outbursts have been detected by  \swift/BAT (15-50 keV)  in 2009 and 2016 as reported by Corbet et al. (2017), although no  information was provided about their  duration and energetics.  To this aim,  we derived the  \swift/BAT count rates at the peak of both events following Corbet et al. (2017). Then we used them with WEBPIMMS in order to estimate the fluxes and corresponding luminosities in the energy band 18-60 keV to allow a proper comparison with the   outburst detected by \int\ (see Table 5). We assumed the same  spectral model and parameter values (i.e. power law)  of the  \int\  outburst detection reported in this work. As for the outburst duration, from the \swift/BAT daily light curve we  inferred  values of  $\sim$18 days and  $\sim$20 days, respectively.  We note that the  X-ray characteristics of all outbursts could be indicative of  classical Type I outbursts from BeXRB, hence reinforcing the proposed nature. We calculated a source duty cycle of $\sim$1.6\% from \int\  observations (18-60 keV) covering the period Jan 2003 -- Sep 2018. In a similar energy band (15-50 keV), we obtained a value of  $\sim$0.8\% from \swift/BAT observations covering a significantly longer period from Feb 2005 to Apr 2023.

Outside X-ray outburst activity, to date \src\ was observed 6 times with sensitive X-ray satellites below 10 keV (5 out of 6 observations are reported for the first time in this work).  Notably, over those 6 X-ray observations spanning about 18 years, \src\ was always detected with a  very similar low X-ray luminosity value in the range  (1-4)$\times$10$^{34}$ erg s$^{-1}$.  Clearly, when \src\ has been observed with sufficient exposure (i.e. $\ge$ 0.5 ks) with sensitive X-ray satellites below 10 keV, it has  been always detected as a persistent low X-ray luminosity source with  limited variability (a factor as high as  4). Above 20 keV, the X-ray luminosity upper limit is $<$3$\times$10$^{34}$ erg s$^{-1}$, as we inferred from deep INTEGRAL observations (16 Ms).  At energies $>$ 20 keV the level of the source persistent low luminosity  is below the sensitivity threshold of current hard 
X-ray missions like \int\ or \swift/BAT.

We note that the above newly reported  X-ray characteristics of \src\  strongly resemble those of persistent low luminosity BeXRBs,  which are a small and rare subgroup  of peculiar BeXRBs (Pfahl et al. 2002). To date, only an handful of such objects have been reported in the literature, amongst the $\sim$100 known classical transient BeXRBs. Persistent low luminosity BeXRBs are mainly characterized by i) long spin period (greater than several tens of seconds), ii) persistent low X-ray luminosity in the range 10$^{34}$--10$^{35}$ erg  s$^{-1}$,  iii)  limited  X-ray  variability (rare and unpredictable increases in flux by a factor of $\le$10). Such properties 
suggest that the compact object orbits the donor Be star in a wide (orbital periods longer than $\sim$ 30 days) 
and nearly circular  (e$<$0.2) orbit, rarely or never crossing the decretion disk star and continuously accreting material from the lower density outer regions of the stellar wind.

\begin{figure}
 \includegraphics[width=6.1cm, angle=270]{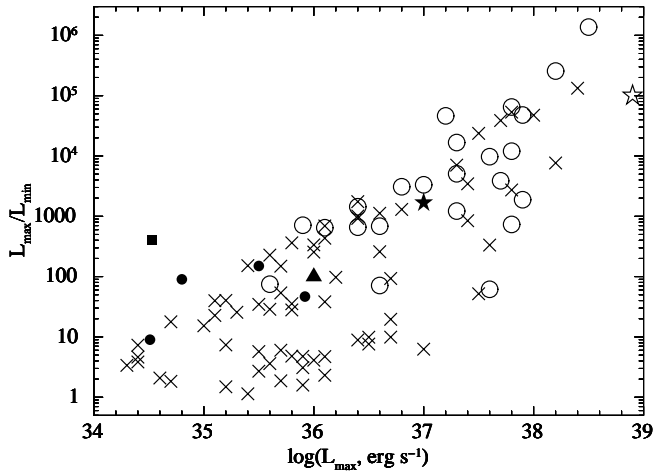} 
 \caption{Source X-ray variability as a function of the maximum luminosity, for XTE J1906+090 (filled triangle), RX J0440.9+4431 (filled star), SAX J0635.2+0533 (filled square), A0538-66 (open star) and various classes of BeXRBs: persistent (filled circles), MW transients (open circles) and SMC transients (crosses).}\label{fig:allbex}
\end{figure}

Our proposed classification of persistent low luminosity BeXRB  for XTE J1906+090 is not at odds with the four X-ray outbursts observed to date. In fact, we note that a similar sporadic variability was already observed  in all the BeXRBs which are traditionally considered as persistent sources, such as RX J0146.9+6121 (La Palombara \& Mereghetti 2006), X Persei (La Palombara \& Mereghetti 2007), and RX J1037.5-5647 (La Palombara et al. 2009). Although these sources are usually observed at a X-ray luminosity level of about 10$^{34}$ erg  s$^{-1}$, occasionally all of them have revealed a luminosity increase up to a few 10$^{35}$ erg  s$^{-1}$, thus with a variability higher than one order of magnitude.

Even more interesting is the case of the BeXRB RX J0440.9+4431, which is the fourth member of the class of persistent BeXRBs originally identified by Reig \& Roche (1999). Between 2010 and 2011 three outbursts were observed for this source, which reached a peak luminosity of about  10$^{36}$ erg  
s$^{-1}$ (La Palombara et al. 2012). These were the only flux increases larger than one order of magnitude observed for this source over a time span of about 25 years. But, very surprisingly, at the end of 2022 this source has shown a new outstanding outburst, reaching a peak X-ray luminosity of about  10$^{37}$ 
erg  s$^{-1}$ (Coley et al. 2023). The origin of these rare and unpredictable events is not clear. Very likely, these large outbursts could be explained by particularly rare episodes of structural changes in the decretion disk of the Be star which could result in a significant enlargement of the disk itself, allowing its crossing with the neutron star compact object. We note that the flux variability of RX J0440.9+4431 is remarkably similar to that observed in XTE J1906+090. In fact, also for this source  four different outbursts were detected in over 20 years of observations; for all of them the peak luminosity was about  10$^{36}$ erg  s$^{-1}$. Apart from these outbursts, similarly to the previous cited sources, XTE J1906+090 was always detected at a X-ray luminosity level of a few 10$^{34}$ erg  s$^{-1}$, never going to quiescence. The occurrence of these sporadic outbursts in these otherwise steady sources suggests that the variability requirement Lmax/Lmin $<$ 10, originally proposed by Reig \& Roche (1999), is no longer valid in order to identify persistent BeXRBs.

The similarity between XTE J1906+090 and the other persistent BeXRBs is shown in Fig 6, where we report the source variability (Lmax/Lmin) as a function of the maximum observed luminosity (Lmax) for various classes of BeXRBs. There RX J0440.9+4431 is represented as a filled star, while the four filled circles represent the other three persistent and low-luminosity BeXRBs, plus CXOU J225355.1+624336, which is a candidate member of this group (La Palombara et al. 2021). On the other hand, open circles and crosses represent, respectively, the transient MW sources (Tsygankov et al. 2017b) and the SMC sources reported by Haberl \& Sturm (2016). For completeness, we report also the two peculiar sources SAX J0635.2+0533 (filled square; La Palombara \& Mereghetti 2017) and LMC 0538-66 (open star; Skinner et al. 1982, Kretschmar et al. 2004). In this figure XTE J1906+090 is reported as a filled triangle. The Lmax/Lmin ratio of this source is fully consistent with that observed in most of the persistent BeXRBs, thus supporting its classification as an additional member of this class of sources.

Corbet et al. (2017) reported the discovery of two periodicities from XTE J1906+090 particularly long (81 days and 173 days) which could be interpreted in terms of orbital and super-orbital period, respectively. This finding supports the possibility of a significantly wide orbit for the source, similarly to the case of persistent low luminosity BeXRBs.

From the spectral point of view, it is interesting to note that the \chandra\ spectrum of XTE J1906+090 (G{\"o}{\v{g}}{\"u}{\c{s} et al. 2005) has been described by a hot black body model (kT $\sim$ 1.5 keV). In fact, this type of thermal component has been observed in several low-luminosity and long-period HMXBs and, in particular, in the four confirmed persistent BeXRBs (La Palombara et al. 2012). In all cases the black body radius is remarkably consistent with the estimated size of the neutron star polar cap (assuming standard values for the neutron star parameters). This finding supports the hypothesis that the observed thermal component originates at the surface of the neutron star.

It would be very interesting to investigate the presence and the properties of this component also in XTE J1906+090, but the available exposure time and statistics of the current soft X-ray data (below 10 keV) prevent us from a more detailed spectral investigation. Longer soft X-ray observations (e.g. with \xmm) would be particularly useful to perform a much deeper X-ray spectral investigation of the additional thermal spectral component, in order to support our proposed scenario of a persistent low luminosity BeXRB. Still, the final word on the real nature of XTE J1906+090 is up to near-infrared or optical spectroscopy of the putative counterpart, which is still lacking.

\section*{Acknowledgements}
We warmly thank  the referee for the prompt and constructive report. 
This research has made use of data and software provided by the High Energy Astrophysics Science Archive Research Center (HEASARC), which is a service of the Astrophysics Science Division at 
NASA/GSFC. This work has made use of the Swift/BAT transient monitor results
provided by the Swift/BAT team: \url{http://swift.gsfc.nasa.gov/docs/swift/results/transients/}. 
This work has made use of data from the ESA mission
{\it Gaia} (\url{https://www.cosmos.esa.int/gaia}), processed by the {\it Gaia}
Data Processing and Analysis Consortium (DPAC,
\url{https://www.cosmos.esa.int/web/gaia/dpac/consortium}).

\section*{Data Availability}
This research has made use of data and software provided by the High Energy Astrophysics Science Archive Research Center (HEASARC), which is a service of the Astrophysics Science Division at 
NASA/GSFC. We used archival data of INTEGRAL, Swift and XMM-Newton observatories.
INTEGRAL data can be retrieved from the INTEGRAL Science Data Centre at \url{http://isdc.unige.ch/integral/archive#Browse}.
XMM-Newton data are publicly accessible  by means of the ESA archive at the link \url{http://nxsa.esac.esa.int/nxsa-web/#search}.
Swift/XRT data are publicly available from the UK Swift Science Data Centre at   \url{https://www.swift.ac.uk/swift_portal/}.  This work has made use of the Swift/BAT transient monitor results
provided by the Swift/BAT team: \url{http://swift.gsfc.nasa.gov/docs/swift/results/transients/}.











\bsp	
\label{lastpage}
\end{document}